\documentclass[showpacs,superscriptaddress,amsmath,amssymb,aps,preprint]{revtex4}
\usepackage{graphicx}
\usepackage{bm}
\usepackage[utf8]{inputenc}
\usepackage{lmodern}
\usepackage{graphicx}
\begin{document}
\title{Negative energy waves and quantum relativistic Buneman instabilities} 
\author{F. Haas}
\affiliation{Departamento de F{\'i}sica, Universidade Federal do Paran\'a, 81531-990, Curitiba, Paran\'a, Brazil}
\author{B. Eliasson}
\affiliation{International Centre for Advanced Studies in Physical Sciences and Institute for Theoretical
Physics, Faculty of Physics \& Astronomy, Ruhr University Bochum, D-44780 Bochum, Germany}
\author{P. K. Shukla}
\affiliation{International Centre for Advanced Studies in Physical Sciences and Institute for Theoretical
Physics, Faculty of Physics \& Astronomy, Ruhr University Bochum, D-44780 Bochum, Germany}
\affiliation{Department of Mechanical and Aerospace Engineering \& Center for Energy Research, University of California
San Diego, La Jolla, CA 92093, U. S. A.}
\revised{6 June 2012}
\begin{abstract}
The quantum relativistic Buneman instability is investigated theoretically using a collective
Klein-Gordon model for the electrons and a cold fluid model for the ions. The growth rate
and unstable wave spectrum is investigated in different parameter regimes corresponding to
various degrees of relativistic and quantum effects. The results may be important for
streaming instabilities involving ion dynamics in very dense plasmas. 
\end{abstract}
\maketitle
\section{Introduction}
The problem of the stability of electron beams propagating through 
a plasma is important in the context of laboratory beam-plasma experiments \cite{Thode73},
inertial fusion schemes \cite{Tabak}, in the solar corona \cite{Klassen03}, astrophysical objects \cite{Zensus97,Piran04}, etc. 
It has been suggested that pulsar glitches are due to a streaming instability \cite{Anderson03,Anderson04}, where super-fluid neutrons 
and superconducting protons co-exist with relativistic electrons \cite{Samuelson10}.
It was early recognized that electron beams propagating
through the plasma can give rise to high-frequency electron plasma waves, or low frequency
ion acoustic waves. The ion motion becomes important if the electrons drift as a whole
through the plasma and typically becomes unstable if the relative drift speed between the electrons
and ions is larger than the ion acoustic speed, as in the original 
Buneman instability \cite{Buneman58} in an unmagnetized plasma (or electrons streaming along the magnetic field lines) 
and the Farley-Buneman instability in magnetized plasmas \cite{Buneman63}.

The physics of the linear Buneman instability is fairly well understood \cite{Briggs} in the non-quantum ($\hbar = 0$) and 
non-relativistic ($1/c = 0$) cases ($\hbar$ is the reduced Planck's constant and $c$ is
the speed of light in vacuum).  
In other situations, relatively thin beams of electrons penetrate a plasma consisting of 
both electrons and ions, giving rise to electron two-stream and ion Buneman instabilities
in different parameter regimes in the non-relativistic \cite{Bret11} and relativistic \cite{Bludman60,Bret08,Bret10} regimes.
An electron beam propagating through
a plasma can give rise to electrostatic two-stream instabilities 
when the wave vector and electric field are aligned, and electromagnetic instabilities
obliquely to the beam direction. 
A typical condition for the ion motion to become important in a current-neutral plasma is 
$\gamma \geq \alpha M/(m Z_i)$, with $\gamma$ being the electron beam gamma factor, $\alpha$ the
beam to plasma density ratio, $Z_i$ the ion charge state, and $M, m$ the ion and electron masses \cite{Bret11}.
To incorporate quantum effects, collective multi-stream Schr\"odinger \cite{Haas00} and Klein-Gordon \cite{Haas12} 
models have been used to investigate the electron quantum two-stream instability for non-relativistic and relativistic
cases. In the non-relativistic regime, the quantum two-stream instability can be physically understood in terms of a free 
energy available due to negative energy modes \cite{brn}. In addition, low frequency linear and nonlinear waves in 
collisional, non-relativistic quantum plasmas can be derived using multi-stream, carrier-envelope methods \cite{br}.

In this work, we consider the quantum relativistic Buneman instability, in which the bulk electrons 
stream against the ions, initially at rest. We compare the classical relativistic case with the
non-relativistic and quantum case, as well as with the combined quantum relativistic case. As a model, we use
a collective Klein-Gordon model \cite{Taka53} for the electrons and a cold plasma model for the ions. 

The paper is organized as follows. In Section II, the pertinent dielectric function and dispersion relation are deduced, starting from the basic hydrodynamic equations for the two-species plasma. In Section III, assuming low frequency and a beam-plasma resonance condition, the corresponding Buneman instability is derived, in the case of negligible quantum effects. Moreover, the instability is intuitively understood in terms of positive and ne\-ga\-ti\-ve energy modes. Section IV and V extend the treatment to the quantum non-relativistic and quantum relativistic cases, respectively. Section VI contains the final remarks and a discussion on the validity conditions of the model.

\section{Dielectric function}
The mathematical model is based on a collective Klein-Gordon equation for the electrons that is cast in to a set of fluid-like equations,
and a cold fluid model for the ions. The electrons are described by a Klein-Gordon field $\psi = R\exp[iS/\hbar]$, so that $R = R(x,t)$ and 
$S = S(x,t)$ can be viewed as, respectively, the amplitude and phase of the collective electron wave function. From the Klein-Gordon equation, the evolution equations for $R$ and $S$ are
obtained as \cite{Taka53,Haas12}
\begin{align}
\label{z1}
R\left(\frac{1}{c^2}\frac{\partial^2}{\partial t^2} - \frac{\partial^2}{\partial x^2}\right)S &- \frac{e R}{c^2}\frac{\partial\phi}{\partial t} + \frac{2}{c^2}\,\frac{\partial R}{\partial t}\left(\frac{\partial S}{\partial t}-e\phi\right) - 2\frac{\partial R}{\partial x}\frac{\partial S}{\partial x}  = 0 \,,\\
\label{z2}
\frac{1}{c^2}\left(\frac{\partial S}{\partial t} - e\phi\right)^2 &- \left(\frac{\partial S}{\partial x}\right)^2 - m^2 c^2 = \frac{\hbar^2}{R}\left(\frac{1}{c^2}\frac{\partial^2}{\partial t^2} - \frac{\partial^2}{\partial x^2}\right)R \,,
\end{align}
where  $-e$ and $m$ are the electron charge and mass, $c$ is the speed of light in vacuum, and $\hbar$ is Planck's constant divided by $2\pi$.  
Equation (\ref{z1}) plays the role of a quantum relativistic electron continuity equation, while Eq. (\ref{z2}) is a relativistic Hamilton-Jacobi equation with a quantum Bohm-like $\propto \hbar^2$ correction term, for the electrons.
The ion continuity and momentum equations for the ion fluid density $n_i$ and velocity $v_i$ are
\begin{align}
\label{z3}
\frac{\partial n_i}{\partial t} + \frac{\partial}{\partial x}(n_i u_i) &= 0 \,, \\
\label{z4}
\frac{\partial u_i}{\partial t} + u_i \frac{\partial u_i}{\partial x} &= - \frac{e}{M}\frac{\partial\phi}{\partial x} 
\,,
\end{align}
respectively, where $M$ is the ion mass. Due to their larger mass, the ions are taken as non-relativistic and non-quantum. 
Also, as a first approximation, no thermal effects are included for either electrons and ions. 
The electrostatic potential $\phi$ is obtained from Poisson's equation
\begin{align}
\label{z5}
\frac{\partial^2 \phi}{\partial x^2} &= - \frac{e}{\varepsilon_0}\left[\frac{R^2}{mc^2}\left(\frac{\partial S}{\partial t}-e\phi\right) + n_i\right] \,.
\end{align}
where $\varepsilon_0$ the vacuum electric permittivity, in terms of the appropriate electron charge density, 
as discussed in more detail in Ref. \cite{Haas12}. Since the interest here is on electrostatic waves only, it is sufficient to write 
the model in terms of $1+1$ dimensions, with a scalar potential $\phi = \phi(x,t)$. 

Assuming the electron fluid streaming as a whole through the ionic fluid, an equilibrium solution is
\begin{equation}
\label{e1}
R = \sqrt{\frac{n_0}{\gamma}} \,, \quad S = - \gamma mc^2 t + p\,x \,, \quad
n_i = n_0 \,, \quad u_i = 0 \,, \quad \phi = 0 \,,
\end{equation}
in the reference frame where ions are at rest, with the relativistic gamma factor $\gamma = [1 + p^2/(m^2 c^2)]^{1/2}$ together 
with $n_0$ as a the equilibrium density. Notice the modified equilibrium electron fluid density, due to the spatial contraction ($\gamma > 1$).

Linearizing the model around the equilibrium (\ref{e1}) with plane wave perturbations $\propto \exp[i(kx-\omega t)]$, one obtains the dielectric function
\begin{equation}
\label{e2}
\varepsilon = 1 - \frac{\omega_{i}^2}{\omega^2} - \frac{\omega_{e}^2}{\gamma}\,\frac{\left[1 - \hbar^2(\omega^2 - c^2 k^2)/(4m^2 c^4)\right]}{\left[\gamma^2(\omega-kv)^2-\hbar^2(\omega^2-c^2 k^2)^2/(4m^2 c^4)\right]} \,,
\end{equation}
the dispersion relation being $\varepsilon = 0$. In Eq. (\ref{e2}), $\omega_e = [n_0 e^2/(m\varepsilon_0)]^{1/2}$ and $\omega_i = [n_0 e^2/(M\varepsilon_0)]^{1/2}$ are the electron and ion plasma 
frequencies, and $v = p/(\gamma m)$ is the beam velocity.

It is instructive to consider the consequences of the dielectric function (\ref{e2}) for the three separate cases (a) relativistic non-quantum; (b) non-relativistic quantum and (c) joint relativistic and quantum. 

\section{Relativistic non-quantum case}
\begin{figure}[!ht]
\centering{\includegraphics[width=10.0cm]{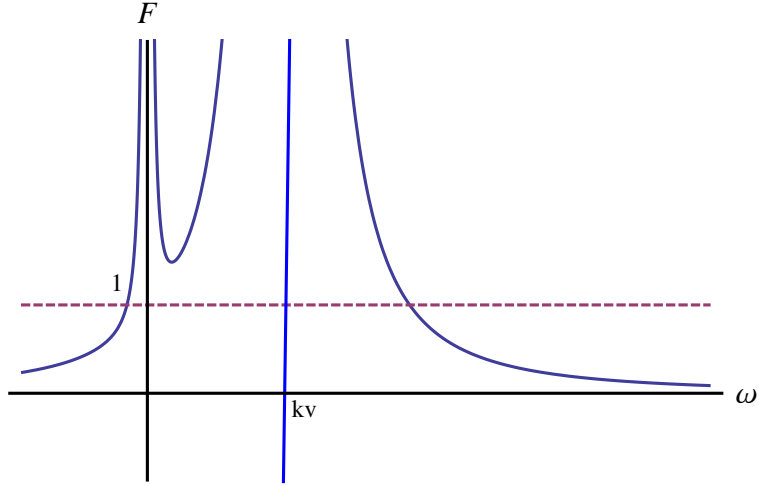}
\caption{Characteristic function $F = F(\omega,k)$ from Eq. (\ref{e3}) in the non-quantum case, showing a generic unstable equilibrium when $F_{\rm min} > 1$}
\label{fig1}}
\end{figure}

Setting formally $\hbar = 0$ in Eq. (\ref{e2}) yield 
\begin{equation}
\label{e3}
\varepsilon = 1 - \frac{\omega_{i}^2}{\omega^2} - \frac{\omega_{e}^2}{\gamma^3(\omega-kv)^2} = 1 - F(\omega,k) \,,
\end{equation}
which also defines a non-quantum characteristic function $F = F(\omega,k)$. The dispersion relation $F = 1$ is a fourth degree polynomial in the wave frequency $\omega$. From the graphics of the characteristic function, it follows that stability (four real solutions) is obtained when the minimum value $F_{\rm min} < 1$. When $F_{\rm min} > 1$ one has two real and two complex conjugate roots, one of them an unstable mode. Finally, $F_{\rm min} = 1$ is marginally stable. In Fig. (\ref{fig1}) an unstable case is depicted. 

Supposing $\gamma(m/M)^{1/3} \ll 1$, which in hydrogen plasma is reasonable except in extreme relativistic situations ($\gamma > 12.3$ or $\beta = v/c > 0.997$), it is straightforward to obtain 
\begin{equation}
\omega_{\rm min} = \left(\frac{m}{M}\right)^{1/3}\gamma k v 
\end{equation}
as the wave frequency for the minimum of the characteristic function. Correspondingly, 
\begin{equation}
\label{c}
F_{\rm min} = \frac{\omega_{e}^2}{\gamma^3 k^2 v^2}\left[1 + \gamma\left(\frac{m}{M}\right)^{1/3}\right] > 1
\end{equation}
for instability. Hence, relativistic effects ($\gamma > 1$) tend to shrink the range of unstable wave numbers. 

To obtain the interpretation of the stability analysis, we first rewrite the dispersion relation as  
\begin{equation}
\label{e4}
(\omega - \omega_n)\,\omega^2 = \frac{\omega_{i}^2 (\omega - k v)^2}{\omega - \omega_p} \,,
\end{equation}
where 
\begin{equation}
\label{pem}
\omega_p = k v + \frac{\omega_e}{\gamma^{3/2}} \,, \quad \omega_n = k v - \frac{\omega_e}{\gamma^{3/2}}
\end{equation}
are relativistic versions of the usual Doppler-shifted beam modes. Notice that when ions are taken as immobile ($M \rightarrow \infty$) one has $\omega_{p,n}$ as exact normal modes. 

It is interesting to focus on resonant wave numbers and low frequency modes such that 
\begin{equation}
\label{k}
k v \simeq \frac{\omega_e}{\gamma^{3/2}} \,, \quad \omega \ll k v \,,
\end{equation}
implying $\omega_n \simeq 0$. One then immediately obtains the estimates
\begin{equation}
\omega - \omega_p \simeq - 2 k v \,, \quad (\omega - k v)^2 \simeq k^2 v^2 \,, \quad \omega - \omega_n \simeq \omega \,,
\end{equation}
so that Eq. (\ref{e4}) becomes
\begin{equation}
\label{e5}
\omega^3 = - \frac{m}{2M}\frac{\omega_{e}^3}{\gamma^{3/2}} \,.
\end{equation}
The reason why the dispersion relation (a fourth degree polynomial equation) can be converted into Eq. (\ref{e5}), 
which is of the third degree, is the assumption (\ref{k}). Far from resonant or high frequency normal modes cannot be detected in this manner.

Equation (\ref{e5}) has one real
\begin{equation}
\omega = - \left(\frac{m}{2M}\right)^{1/3}\frac{\omega_e}{\sqrt{\gamma}}
\end{equation}
and two complex conjugate roots,
\begin{equation}
\label{e6}
\omega = \left(\frac{m}{16M}\right)^{1/3}\frac{\omega_e}{\sqrt{\gamma}}\,(1 \pm i\sqrt{3}) \,.
\end{equation}
The plus sign in Eq. (\ref{e6}) correspond to an exponentially growing mode. Note that the growth rate becomes smaller due to relativistic effects. In addition, for all three modes one get 
\begin{equation}
\frac{\omega}{k v} \sim \gamma\left(\frac{m}{M}\right)^{1/3} \,,
\end{equation}
apart from numerical factors of order unity, so that the low frequency assumption is acceptable with the exception of extremely relativistic beams (in the case of hydrogen plasma where $m/M = 1/1867$). Moreover, the instability condition (\ref{c}) is also fulfilled by the resonant wave numbers in Eq. (\ref{k}). 

The time-averaged energy density $<\!\!W\!\!>$ in a dielectric medium can be shown \cite{Landau} to be given by 
\begin{equation}
\label{w}
<\!\!W\!\!> = \frac{\varepsilon_0 |\delta E|^2}{4}\frac{d}{d\omega}\,\left[\omega \varepsilon_{r}(\omega)\right]_{\omega_r} \,,
\end{equation}
where $\delta E$ is the electric field perturbation and $\varepsilon_{r}, \omega_r$ are the real parts of the dielectric function and wave frequency. In the present dissipation-free problem, $\varepsilon = \varepsilon_r$. In Eq. (\ref{w}), both field and particle energy contributions are already taken into account. 

From the above well-known result, it is found that the energy content of the different modes depends on the quantity 
\begin{equation}
\omega\frac{\partial\varepsilon}{\partial\omega} = 2\frac{\omega_{i}^2}{\omega^2} + \frac{2\omega_{e}^2 \omega}{\gamma^3(\omega - k v)^3} \,,
\end{equation}
calculated from Eq. (\ref{e3}) considering $\omega = \omega_r$. By inspection, the ion contribution has always a positive energy. Concerning electrons, if the wave velocity is larger than the beam's mode the energy is positive. From Eq. (\ref{pem}) one has that $\omega = \omega_p$ is a positive energy mode, while $\omega = \omega_n$ is a negative energy mode (without loss of generality, we use in this work the convention $\omega > 0, k > 0$). One can understand the result as follows. When the Doppler-shifted frequency of the beam mode is negative, its energy is also negative. Moving through the plasma, the beam is then slowed down, losing energy which is the seed for the growing amplitude of the wave. 

In conclusion, in this Section the low-frequency instability due to a negative energy relativistic beam mode in an oscillating ionic background was investigated. Relativistic effects tend to produce a smaller range of unstable wave numbers, as well as a smaller growth rate, in comparison to the $\gamma \simeq 1$ case.

\section{Non-relativistic quantum case}
\begin{figure}[!ht]
\centering{\includegraphics[width=10.0cm]{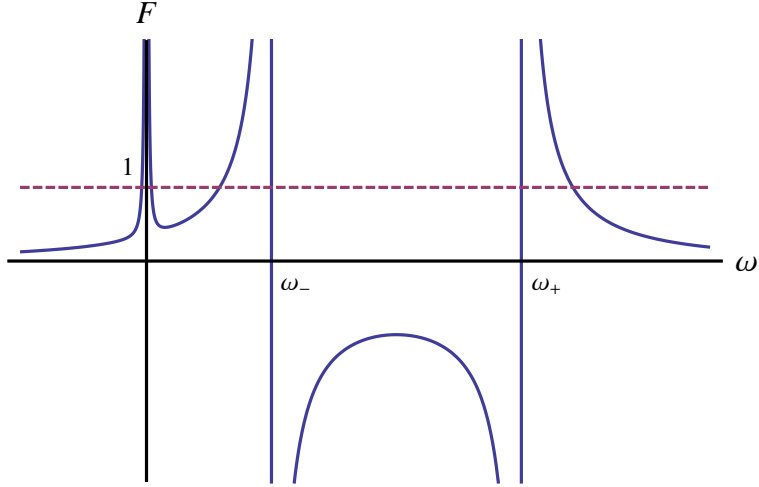}
\caption{Characteristic function $F = F(\omega,k)$ from Eq. (\ref{e7}) in the non-relativistic case, showing a generic stable equilibrium when $F_{\rm min} < 1$. In the graphic, $\omega_{\pm} = k v \pm \hbar k^2/(2 m)$}
\label{fig2}}
\end{figure}

Setting formally $1/c = 0$ in Eq. (\ref{e2}) yields
\begin{equation}
\label{e7}
\varepsilon = 1 - \frac{\omega_{i}^2}{\omega^2} - \frac{\omega_{e}^2}{(\omega-kv)^2 - \hbar^2 k^4/(4m^2)} = 1 - F(\omega,k) \,,
\end{equation}
which also defines a non-relativistic characteristic function $F = F(\omega,k)$. The dispersion relation $F = 1$ is again a fourth degree polynomial in the wave frequency $\omega$. However, there are qualitative changes in comparison with the non-quantum case, as apparent in Fig. (\ref{fig2}) drawn for $v > \hbar k/(2m)$. It can be verified that the maximum of the characteristic function for frequencies $k v - \hbar k^2/(2m) < \omega < k v +  \hbar k^2/(2m)$ is always negative. Hence one need $F_{\rm min} > 1$ for instability, where the corresponding wave frequency $\omega_{\rm min}$ satisfy $0 < \omega_{\rm min} < k v -  \hbar k^2/(2m)$. In passing, the case $v \leq \hbar k/(2m)$ can be shown to produce only linearly stable oscillations and will be not considered.

The dispersion relation $F = 1$ can be rewritten as 
\begin{equation}
(\omega - \omega_n)\,\omega^2 = \frac{\omega_{i}^2 \left[(\omega - k v)^2 - \hbar^2 k^4/(4 m^2)\right]}{\omega - \omega_p} \,,
\end{equation}
where 
\begin{equation}
\label{pn}
\omega_p = k v + \left(\omega_{e}^2 + \frac{\hbar^2 k^4}{4 m^2}\right)^{1/2} \,, \quad \omega_n = k v - \left(\omega_{e}^2 + \frac{\hbar^2 k^4}{4 m^2}\right)^{1/2} 
\end{equation}
are quantum versions of the usual Doppler-shifted beam modes. Notice that when ions are taken as immobile ($M \rightarrow \infty$) one has $\omega_{p,n}$ as exact normal modes. 

In analogy with the previous Section, it is interesting to focus on wave numbers such that $\omega_n \simeq 0$ and on low frequencies, or
\begin{equation}
k v \simeq \left(\omega_{e}^2 + \frac{\hbar^2 k^4}{4 m^2}\right)^{1/2} \,, \quad \omega \ll k v \,.
\end{equation}

Proceeding as before, the result is
\begin{equation}
\label{t}
\omega^3 = - \frac{m \omega_{e}^3}{2 M}\, \theta_{\pm}(H) \,,
\end{equation}
where 
\begin{equation}
\label{h}
H = \frac{\hbar\omega_e}{m v^2}
\end{equation}
measures the strength of quantum effects and where it was defined
\begin{equation}
\label{tts}
\theta_{\pm}(H) = \frac{H}{\sqrt{2}}\left[1 \pm (1 - H^2)^{1/2}\right]^{-1/2} \,.
\end{equation}
One should compare the relativistic and quantum expressions for the low frequency beam modes in Eqs. (\ref{e5}) and (\ref{t}),
respectively. In addition, $H$ is bigger for dense plasmas since $\omega_e \propto n_{0}^{1/2}$.

\begin{figure}[!ht]
\centering{\includegraphics[width=10.0cm]{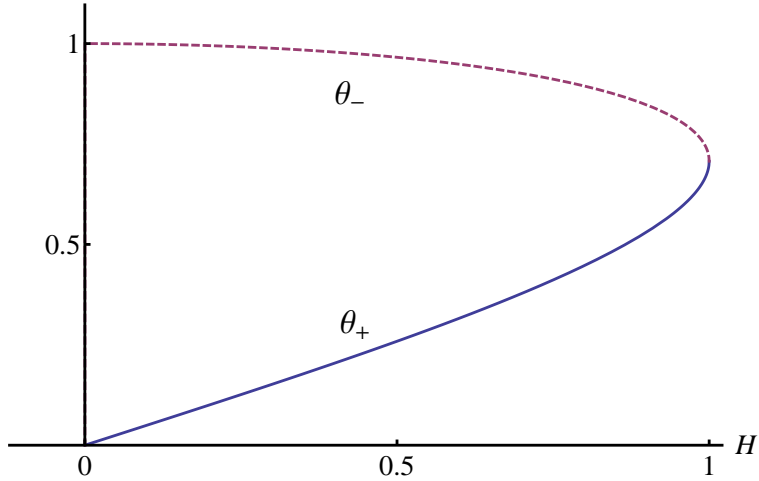}
\caption{Functions $\theta_{\pm}$ from Eq. (\ref{tts}). Bottom, line: $\theta_+$. Upper, dashed: $\theta_-$}
\label{fig3}}
\end{figure}

The resonance condition can be explicitly solved yielding 
\begin{equation}
\label{kk}
k v = \frac{\omega_e}{\theta_{\pm}(H)} \,,
\end{equation}
which gives real wave numbers provided $H^2 \leq 1$. In particular, the previous resonant wave number is now split in two, 
due to the quantum recoil term $\hbar^2 k^4/(4 m^2)$ in Eq. (\ref{e7}). Moreover, one has 
\begin{equation}
\label{xx}
\theta_{+}(H) = \frac{H}{2}\left(1 + \frac{H^2}{8}\right) + O(H^5) \,, \quad \theta_{-}(H) = 1 - \frac{H^2}{8} + O(H^4) \,,
\end{equation}
so that $\theta_{\pm}$ have, respectively, a purely quantum and semiclassic nature. In Fig. (\ref{fig3}) the behavior $\theta_{\pm}(H)$ is shown. The two modes coalesce when $H = 1$.

Similarly to the non-quantum case, Eq. (\ref{t}) can be solved yielding purely oscillatory, 
\begin{equation}
\label{q1}
\omega = - \left(\frac{m\,\theta_{\pm}(H)}{2M}\right)^{1/3}\omega_e \,,
\end{equation}
damped,
\begin{equation}
\label{q2}
\omega = \left(\frac{m\,\theta_{\pm}(H)}{16M}\right)^{1/3}\omega_{e}(1 - i\sqrt{3}) \,,
\end{equation}
and exponentially growing
\begin{equation}
\label{q}
\omega = \left(\frac{m\,\theta_{\pm}(H)}{16M}\right)^{1/3}\omega_{e}(1 + i\sqrt{3}) 
\end{equation}
waves. Now one has not only one, but two unstable low frequency modes, according to the plus or minus sign chosen in $\theta_{\pm}$. However, since $\theta_{\pm}(H) \leq	1$, it is apparent from Eq. (\ref{q}) that quantum effects are stabilizing. 

For consistency, it remains to check the low frequency assumption. From Eqs. (\ref{q1})--(\ref{q}) it follows that 
\begin{equation}
\frac{\omega}{kv} \sim \left(\frac{m}{M}\right)^{1/3}\theta_{\pm}^{4/3} \ll 1 \,.
\end{equation}
Besides, after some algebra one finds from Eq. (\ref{e7}),
\begin{equation}
\omega_{\rm min} = \left[\frac{m}{M}\,
\theta_{\pm}(H)\right]^{1/3} \quad \Rightarrow \quad F_{\rm min} \simeq 1 + 3\left[\frac{m}{M\theta_{\pm}^{2}(H)}\right]^{1/3} > 1 
\end{equation}
in accordance with the general instability condition. 

Concerning the total energy, one should analyze the quantity
\begin{equation}
\omega\frac{\partial\varepsilon}{\partial\omega} = \frac{2\omega_{i}^2}{\omega^2} + \frac{2\omega_{e}^2\omega(\omega - k v)}{\left[(\omega - k v)^2 - \hbar^2 k^4/(4 m^2)\right]^2} \,.
\end{equation}
By using Eq. (\ref{pn}) one concludes that $\omega=\omega_{p}$ and  $\omega=\omega_{n}$ are positive and negative energy modes, respectively.

\section{Relativistic quantum case}

Having in mind the fruitful results following from the low frequency assumption, in this work the full relativistic and quantum dispersion relation (\ref{e2}) will not be investigated in detail. Rather, it is interesting to restrict to slow waves such that
\begin{equation}
\label{sl}
\omega^2 \ll c^2 k^2 \,,
\end{equation}
which implies
\begin{equation}
\label{rq}
\varepsilon = 1 - \frac{\omega_{i}^2}{\omega^2} - \frac{\omega_{e}^2}{\gamma}\,\frac{\left[1 + \hbar^2 k^2/(4m^2 c^2)\right]}{\left[\gamma^2(\omega-kv)^2-\hbar^2 k^4/(4m^2)\right]} \,.
\end{equation}
The qualitative form of the characteristic function derived from Eq. (\ref{rq}) is similar to the non-relativistic quantum characteristic function from Section IV, with quantitative changes due to $1/c \neq 0, \gamma > 1$. 

Notice that fast wave propagation not satisfying Eq. (\ref{sl}) can be relevant in some ins\-tan\-ces, but is outside the scope of the present work. For instance, the usual Bohm correction is not valid for waves with phase speed greater than the speed of light \cite{Melrose}. Moreover, the simplified dielectric function (\ref{rq}) does not produce the high frequency pair modes \cite{Kowalenko} admitted by the fully relativistic dispersion relation. 

Once again, the dispersion relation $F = 1$ is a fourth degree polynomial in the wave frequency $\omega$, which can be written as
\begin{equation}
(\omega - \omega_n)\,\omega^2 = \frac{\omega_{i}^2 \left[(\omega - k v)^2 - \hbar^2 k^4/(4 \gamma^2 m^2)\right]}{\omega - \omega_p} \,,
\end{equation}
where 
\begin{eqnarray}
\label{wp}
\omega_p &=& k v + \left[\frac{\omega_{e}^2}{\gamma^3}\left(1 + \frac{\hbar^2 k^2}{4 m^2 c^2}\right) + \frac{\hbar^2 k^4}{4 \gamma^2 m^2}\right]^{1/2} \,, \\
\label{wn}
\omega_n &=& k v - \left[\frac{\omega_{e}^2}{\gamma^3}\left(1 + \frac{\hbar^2 k^2}{4 m^2 c^2}\right) + \frac{\hbar^2 k^4}{4 \gamma^2 m^2}\right]^{1/2} \,.
\end{eqnarray}
Neglecting the ion correction, these are exact beam modes for the dielectric function (\ref{rq}). They correspond to Doppler-shifted relativistic modifications of the usual Bohm-Pines dispersion relation \cite{Bohm}. 

Similarly to the non-quantum or non-relativistic cases, it is useful to focus on wave numbers such that $\omega_n \simeq 0$ and on low frequencies, or
\begin{equation}
k v \simeq \left[\frac{\omega_{e}^2}{\gamma^3}\left(1 + \frac{\hbar^2 k^2}{4 m^2 c^2}\right) + \frac{\hbar^2 k^4}{4 \gamma^2 m^2}\right]^{1/2} 
\,, \quad \omega \ll k v \,.
\end{equation}

The resonance condition gives
\begin{equation}
\frac{k v}{\gamma} = \frac{\omega_e}{\varphi_{\pm}(H,\gamma)} \,,
\end{equation}
where
\begin{equation}
\label{qrb}
\varphi_{\pm}(H,\gamma) = \frac{H}{\sqrt{2}}\left[1-\frac{\beta^2 H^2}{4\gamma^3} \pm \left(\left(1-\frac{\beta^2 H^2}{4\gamma^3}\right)^2 - \frac{H^2}{\gamma^5}\right)^{1/2}\right]^{-1/2} \,,
\end{equation}
with $\beta = v/c$ and $H \neq 0$  given by Eq. (\ref{h}). 

Proceeding as before, the result is
\begin{equation}
\label{qb}
\omega^3 = - \frac{m \omega_{e}^3}{2 \gamma^4 M}\, \varphi_{\pm}(H,\gamma)\left(1+\frac{\beta^2\gamma^2 H^2}{4\varphi_{\pm}^{2}(H,\gamma)}\right) \,.
\end{equation}
For $\gamma \rightarrow 1, \beta \rightarrow 0$, one has $\varphi_{\pm}(H,\gamma) \rightarrow \theta_{\pm}(H)$. Moreover, the $\propto \beta^2$ term in Eq. (\ref{qb}) vanishes, so that the quantum non-relativistic results in Eqs. (\ref{t}) and (\ref{kk}) are recovered. On the other hand, choosing $\varphi_{-}(H,\gamma)$ (the semi-classical branch) and then letting $H \rightarrow 0$ it is easy to verify that the relativistic, non-quantum results from Section III are recovered, considering $\varphi_{-}(H,\gamma) = \gamma^{5/2} + O(H^2)$. 

It will be assumed that $\varphi_{\pm}(H,\gamma)$ is not complex, which is true either for
\begin{equation}
\label{bb}
H < \frac{2\gamma}{\beta^2}\left(\sqrt{\gamma} - \frac{1}{\sqrt{\gamma}}\right) = 1 + \beta^2 + \frac{31\beta^4}{32} + O(\beta^6)
\end{equation}
or for 
\begin{equation}
\label{bbb}
H > \frac{2\gamma}{\beta^2}\left(\sqrt{\gamma}+\frac{1}{\sqrt{\gamma}}\right) = \frac{4}{\beta^2} + 2 + \frac{13 \beta^2}{8} + \frac{23\beta^4}{16} + O(\beta^6) \,.
\end{equation}
Equation (\ref{bb}) extends the condition $H < 1$ from the preceding Section to the relativistic domain, while Eq. (\ref{bbb}) seemingly provides a new parameter regime not existent in the non-relativistic approximation. However, in practice it can be verified that the inequality (\ref{bbb}) can be attained only for non-realistic values of $H$ (at least $H > 10.8$). Hence Eq. (\ref{bb}) gives the only important constraint, shown in Fig (\ref{fig4}) below. We note that the relativistic effects allows bigger $H$ values. 

\begin{figure}[!ht]
\centering{\includegraphics[width=8.0cm]{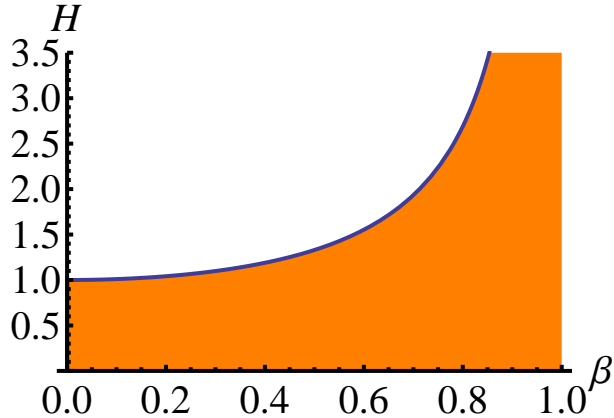}
\caption{The filled area shows the allowable values of the quantum parameter $H$ in terms of $\beta = v/c$, as found from Eq. (\ref{bb})}
\label{fig4}}
\end{figure}

\begin{figure}[!ht]
\centering{\includegraphics[width=14.0cm]{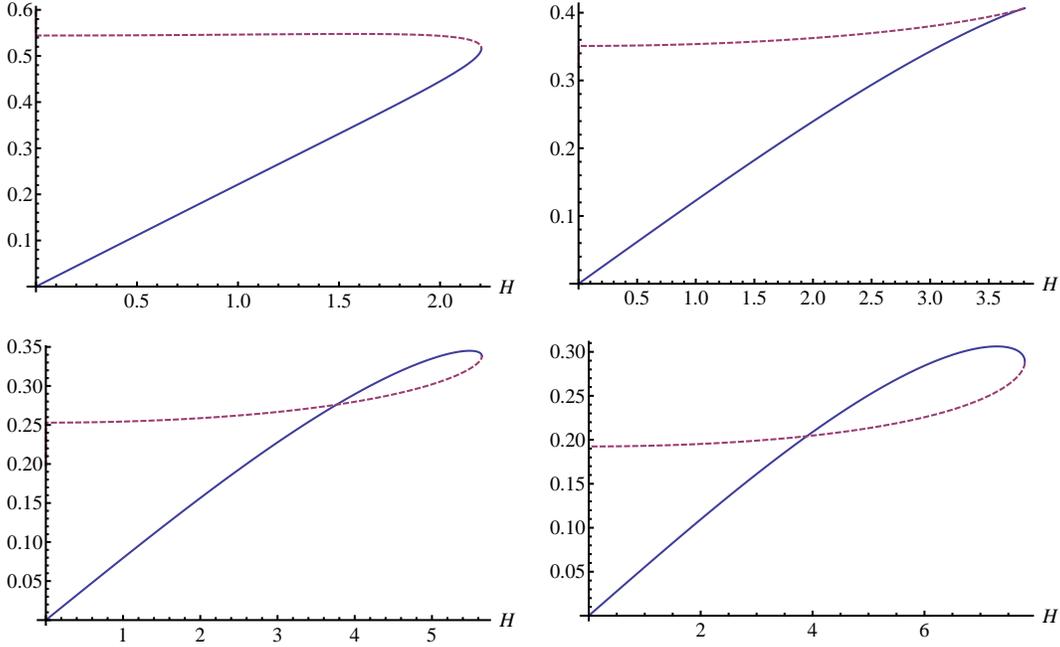}
\caption{Functions $\zeta_{\pm}(H,\gamma)$ from Eq. (\ref{ze}). In all plots, the dashed curve is $\zeta_{-}(H,\gamma)$, while $\zeta_{+}(H,\gamma)$ is the other one. Parameters are: upper, left, $\gamma = 1.5$; upper, right, $\gamma = 2.0$; bottom, left, $\gamma = 2.5$; bottom, right, $\gamma = 3.0$}
\label{fig5}}
\end{figure}

From Eq. (\ref{qb}) it is obvious that the instability analysis from the preceding Section applies, once the replacement $\theta_{\pm}(H) \rightarrow \zeta_{\pm}(H,\gamma)$ is made, where 
\begin{equation}
\label{ze}
\zeta_{\pm}(H,\gamma) \equiv \frac{\varphi_{\pm}(H,\gamma)}{\gamma^4}\left[1+\frac{\beta^2\gamma^2 H^2}{4\varphi_{\pm}^{2}(H,\gamma)}\right] \,.
\end{equation}
Moreover, generalizing Eq. (\ref{xx}) one has 
\begin{equation}
\zeta_{+}(H,\gamma) = \frac{H}{2\gamma^2}\left[1 + \frac{(2-\gamma^2)H^2}{8\gamma^5}\right] + O(H^5) \,, \quad \zeta_{-}(H,\gamma) = \frac{1}{\gamma^{3/2}}\left[1 - \frac{(2-\gamma^2)H^2}{8\gamma^5}\right] + O(H^4) \,,
\end{equation}
so that $\zeta_{\pm}(H,\gamma)$ have, respectively, a purely quantum and semiclassic nature. In Fig. (\ref{fig5}) the behavior of $\zeta_{\pm}(H,\gamma)$ is shown for different values of the relativistic parameter $\gamma$. The two modes coalesce at the maximal quantum parameter found from Eq. (\ref{bb}). Overall, the relativistic effects produce a smaller value of $\zeta_{\pm}(H,\gamma)$ and hence a smaller instability growth rate, which is proportional to $[\zeta_{\pm}(H,\gamma)]^{1/3}$. 

Finally, we note that $\omega=\omega_p$ [Eq. (\ref{wp})] can be shown to be a positive energy mode, 
while $\omega=\omega_n$ [Eq. (\ref{wn})] is a negative energy mode.
\section{Conclusion}

Here we give some notes on the validity of our model. One question in the approach refers to the magnetic field generation due to the unbalanced current at equilibrium. Following the discussion of Bludman {\it et al.} \cite{Bludman60} we can examine the neglect of static magnetic fields according to:
\begin{equation}
\nabla\times{\bf B} = - \mu_0 n_0 e {\bf v} \quad \Rightarrow \quad B \sim \mu_0 n_0 e L v \,,
\end{equation}
where ${\bf B}$ is the magnetic field, $\mu_0$ is the vacuum permeability and $L$ is some characteristic dimension of a beam of finite cross section. From this one can estimate the Larmor frequency $\omega_L$ as
\begin{equation}
\omega_L = \frac{eB}{\gamma m} \sim \frac{n_0 e^2 \mu_0 L v}{\gamma m} = \frac{\omega_{e}^2 L v}{\gamma c^2} \quad \Rightarrow \quad \frac{\omega_L}{\omega_e} = \frac{\beta L \omega_e}{\gamma c} \,.
\end{equation}
For low frequency, the last quantity may be significant, so that the neglect of a return current may be unjustified. However, one can always have $\omega_{L}/\omega_e \ll 1$ for sufficiently small $L$.

Another question is under which ideal experimental conditions the quantum relativistic Buneman instability can be investigated, in particular in view of the quantum effects measured by the parameter $H$ in Eq. (\ref{h}). Using S.I. units one has $H = 7.25 \times 10^{-20} n_{0}^{1/2}/\beta^2$, which attain non-negligible values only for very high densities. For instance, for $\beta = 1/10$ (a weakly relativistic beam) one has $H > 1/10$ only for $n_0 > 1.90 \times 10^{32} m^{-3}$. This corres\-ponds to the parameter regimes of laser-compressed plasma interaction experiments \cite{Azechi91,Kodama01}. Correspondingly, the predicted stabilization effects can be relevant for extremely short wavelengths. 

As a by-product, the present treatment derived the Doppler-shifted relativistic Bohm-Pines dispersion relations in Eqs. (\ref{wp}) and (\ref{wn}). The relativistic effects inhibits dispersion, with a lower group velocity for a given wavelength, as can be verified. This prediction should be tested e.g. with the development of multi-Peta-Watt lasers \cite{Mourou}.

In summary, in this work it was developed an unified methodology for the treatment of low-frequency beam-driven instabilities in relativistic quantum plasmas. A detailed comparison was made considering different parameter regimes (formally, $\hbar = 0$ and $1/c \neq 0$; $\hbar \neq 0$ and $1/c = 0$; both $\hbar \neq 0$ and $1/c \neq 0$). The corresponding Buneman instabilities were shown to be associated to certain beam-plasma negative energy modes. Both quantum and relativistic effects cause a stabilizing effect, with the remark that the quantum effects produce a splitting of the original Buneman mode into two modes, due to the quantum recoil term as discussed in Section IV. Finally, observe that the model equations have certain limitations in particular because of the semiclassical nature (no quantized fields), see \cite{Haas12} for more details. Nevertheless, the dispersion relation (\ref{e2}) can be equally found using more complete, field-theoretic methods \cite{Kowalenko}. In this context, the positive-negative energy waves found here have a more broad applicability, as long as the low frequency assumption is valid.

\begin{acknowledgments}
This work was supported by Conselho Nacional de Desenvolvimento Cient\'{\i}fico e Tecnol\'ogico (CNPq),
as well as by the Deutsche Forschungsgemeinschaft through the project SH21/3-2 of the Research Unit 1048.
\end{acknowledgments}

\end{document}